# Unveiling Topological Hinge States in the Higher-Order Topological Insulator WTe$_2$ Based on the Fractional Josephson Effect

Yong-Bin Choi[1,*], Jinho Park[1,2,*], Woochan Jung[1], Sein Park[1], Mazhar N. Ali[3,4], and Gil-Ho Lee[1,5,†]

[1]Department of Physics, POSTECH, Pohang, Republic Korea

[2]Department of Mechanical Engineering, Columbia University, New York, NY 10027, USA

[3]Kavli Institute of Nanoscience, Delft University of Technology, Delft, Netherlands

[4]Department of Quantum Nanoscience, Delft University of Technology, Delft, Netherlands

[5]Asia Pacific Center for Theoretical Physics, Pohang, Republic of Korea

[*]These authors contributed equally.

[†]Corresponding authors: lghman@postech.ac.kr (G.-H.L.)

**Higher-order topological insulators (HOTIs) represent a novel class of topological materials, characterised by the emergence of topological boundary modes at dimensions two or more lower than those of bulk materials. Recent experimental studies[1-4] have identified conducting channels at the hinges of HOTIs, although their topological nature remains unexplored. In this study, we investigated Shapiro steps in Al–WTe$_2$–Al proximity Josephson junctions (JJs) under microwave irradiation and examined the topological properties of the hinge states in WTe$_2$. Specifically, we analysed the microwave frequency dependence of the absence of the first Shapiro step in hinge-dominated JJs, attributing this phenomenon to the 4$\pi$-periodic current–phase relationship characteristic of topological JJs. These findings may encourage further research into topological superconductivity with topological hinge states in superconducting hybrid devices based on HOTIs. Such advances could lead to the realisation of Majorana zero modes for topological quantum physics and pave the way for applications in spintronic devices.**

Three-dimensional (3D) higher-order topological insulators (HOTIs) represent a novel class of topological materials[1-22], characterised by topologically protected 1D conducting hinge states and topologically trivial and gapped 2D surface and 3D bulk states. Among the distinctive topological features of HOTIs, spin–momentum locking has garnered considerable attention across diverse fields, including condensed matter physics[1-25], photonics[26-30], and electrical circuit design[31]. Specifically, in condensed matter physics, the hinge states of HOTIs are under investigation for potential applications in spintronic devices[3, 12, 21] and superconducting Josephson diodes[22]. Furthermore, topological superconductivity is theorised to originate from interactions between hinge states and *s*-wave superconductors, as proposed in the context of topological materials[32-34], quantum spin Hall systems[35], and semiconductors[36, 37]. However, experimental observations of HOTIs are complicated by the effects of topologically trivial surface and bulk states[1-4, 7, 8, 12-16], which hinder interpretation and limit direct evidence. Nevertheless, numerous investigations into HOTIs and their properties, including studies on pre-patterned Josephson junctions (JJs) with tungsten ditelluride ($WTe_2$)[7-10], switching current distributions in bismuth[4] and $WTe_2$[8], and spin transport in $Cd_3As_2$[12], are underway.

Recent theoretical studies report that $WTe_2$, initially identified as a type-II Weyl semimetal[38], may also exhibit features characteristic of a HOTI[6]. Specifically, a higher-order topological phase can emerge when two adjacent Weyl points merge in momentum space, opening a surface state gap within a crystal structure that preserves mirror symmetry[6]. In the monolayer limit of $WTe_2$, 1D topological edge states, characteristic of a quantum spin Hall insulator, have been observed in transport measurements[39-41]. Similarly, in the multilayer limit of $WTe_2$, previous studies have demonstrated the presence of conducting states at the hinges of $WTe_2$ crystals by analysing the magnetic field interference pattern of the Josephson critical current in $WTe_2$-based JJs and providing a spatial distribution of the Josephson current density[2, 7-9]. This technique has proven to be a powerful tool for investigating surface or hinge states, even in the presence of dominant bulk conduction. Furthermore, scanning tunnelling microscopy-based studies have provided evidence of hinge states at the step edges of $WTe_2$ crystals[42], as well as bismuth crystals and their alloys[1, 14-16, 19, 20]. However, the topological nature of such hinge states is yet to be conclusively verified.

JJs offer a valuable platform for investigating the topological properties of hinge states.

Specifically, when two *s*-wave superconducting leads are coupled by topologically protected conduction channels, the Andreev bound state (ABS) and the corresponding current phase relationship (CPR), $I_s = (2e/\hbar)\, dE_{ABS}/d\varphi$, exhibit $4\pi$ periodicity, unlike the $2\pi$ periodicity observed in topologically trivial JJs[32-37, 43-46]. In the above equation, $\hbar = h/2\pi$ represents the reduced Planck constant, $e$ denotes the electron charge, $\varphi$ represents the macroscopic quantum phase difference between two superconductors, and $E_{ABS}$ denotes the energy of the ABS. The $4\pi$-periodic ABS, linked to the fractional Josephson effect, results from the parity conservation of Majorana zero modes[32-37, 43-54] (Fig. 1a). This phenomenon is experimentally observed through Shapiro step measurements under microwave irradiation. For non-topological JJs with a $2\pi$-periodic ABS, Shapiro steps appear at bias voltages equivalent to integer multiples of $V_0 = hf/2e$, where $f$ denotes the microwave frequency. Conversely, for topological JJs with a $4\pi$-periodic ABS, Shapiro steps occur at bias voltages equivalent to even multiples of $V_0$, resulting in the missing of the first Shapiro step (Fig. 1b). This disappearance of the first Shapiro step has been observed in JJs fabricated from various topological materials, including strong spin–orbit-coupled semiconducting nanowires[47], 2D HgTe quantum wells[48, 49], 3D topological insulators[50-53], and Dirac semimetals[54].

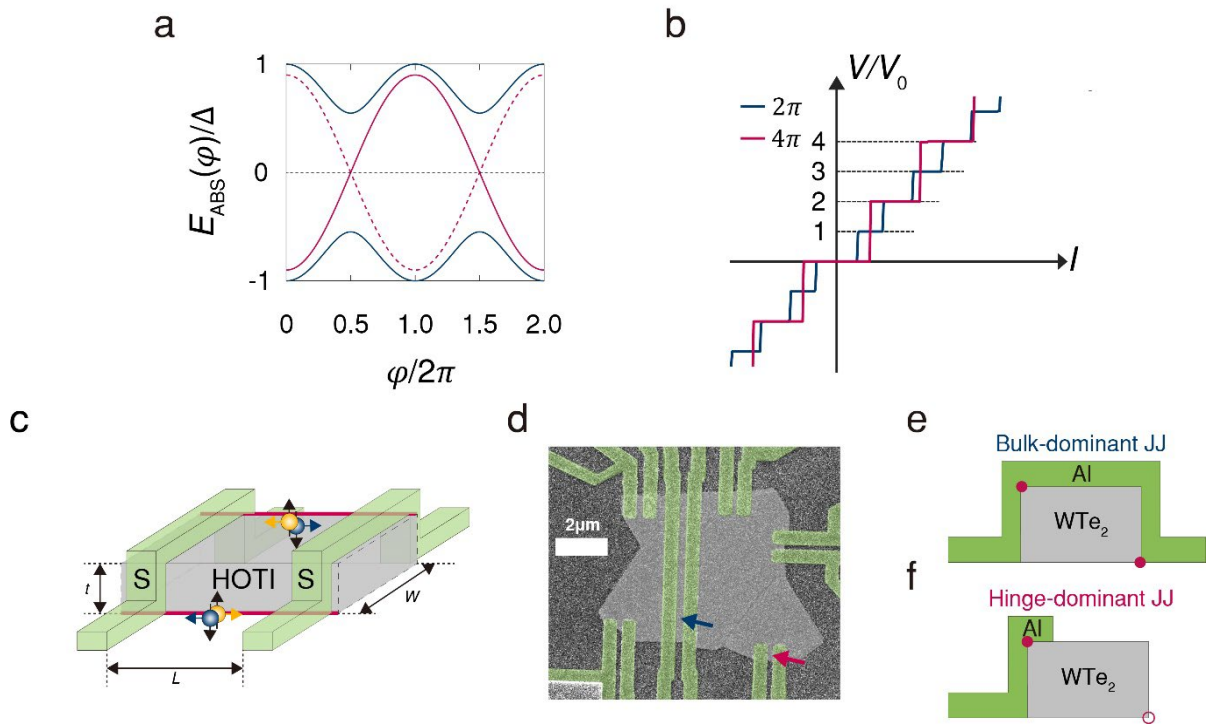

**Fig. 1 Schematic and experimental representations of HOTIs and JJs. a,** Energy spectra of the ABSs, $E_{\mathrm{ABS}}$, as a function of the macroscopic quantum phase difference, $\varphi$, between two superconductors. Here, $\Delta$ represents the superconducting gap. Gapless $4\pi$-periodic ABSs (red solid and dashed lines) and gapped $2\pi$-periodic ABSs (blue solid lines) are illustrated. **b,** Numerically derived current–voltage characteristics depicting Shapiro steps under microwave irradiation for $2\pi$- and $4\pi$-periodic ABSs. The voltage is normalised by $V_0 = hf/2e$, where $h$ denotes Planck constant, $f$ represents the microwave frequency, and $e$ denotes the electron charge. **c,** Schematic of a JJ based on a HOTI. Spin-up (yellow) and spin-down (blue) electrons travel in opposite directions along the topological hinge state (red lines), which is in contact with superconducting (*S*) electrodes (green). *W*, *L*, and *t* denote the junction width, junction length, and thickness of the JJ, respectively. **d,** False-colour scanning electron micrograph of the device. Bulk-dominant JJs (bJJ) and hinge-dominant JJs (hJJ) are indicated by blue and red arrows, respectively. **e, f,** Schematic side views of bJJ and hJJ, respectively. Grey regions represent $WTe_2$ flakes, while red regions indicate hinge states. Filled (solid) and empty (void) red circles depict hinge states coupled and decoupled with aluminium superconducting electrodes, respectively.

In this study, we explored the topological nature of hinge states in multilayer $WTe_2$ by analysing Shapiro steps in $WTe_2$-based JJs. Aluminium superconducting leads were coupled to multilayer $WTe_2$ flakes in two electrode configurations: a bJJ and an hJJ, as depicted in Figs. 1c–e. These configurations allow a comparative investigation of Shapiro steps, highlighting variations in the hinge-to-bulk contribution ratio. The first Shapiro step is missing in the hJJ but not in the bJJ, providing evidence for the topological nature of hinge states in multilayer $WTe_2$. The dependence of this first Shapiro step missing in the hJJ on microwave power and frequency is consistent with the predictions of the resistively and capacitively shunted junction (RCSJ) model, ruling out a topologically trivial origin. Furthermore, the magnitude of the $4\pi$-periodic Josephson current, estimated from Shapiro step measurements, corresponds closely to the Josephson current flowing through the hinge states. This current is derived from the edge-enhancement of the Josephson current density, as determined from magnetic interference patterns[2]. These observations indicate the topological nature of hinge states in multilayer $WTe_2$, establishing a foundation for the study of Majorana physics in layered topological materials.

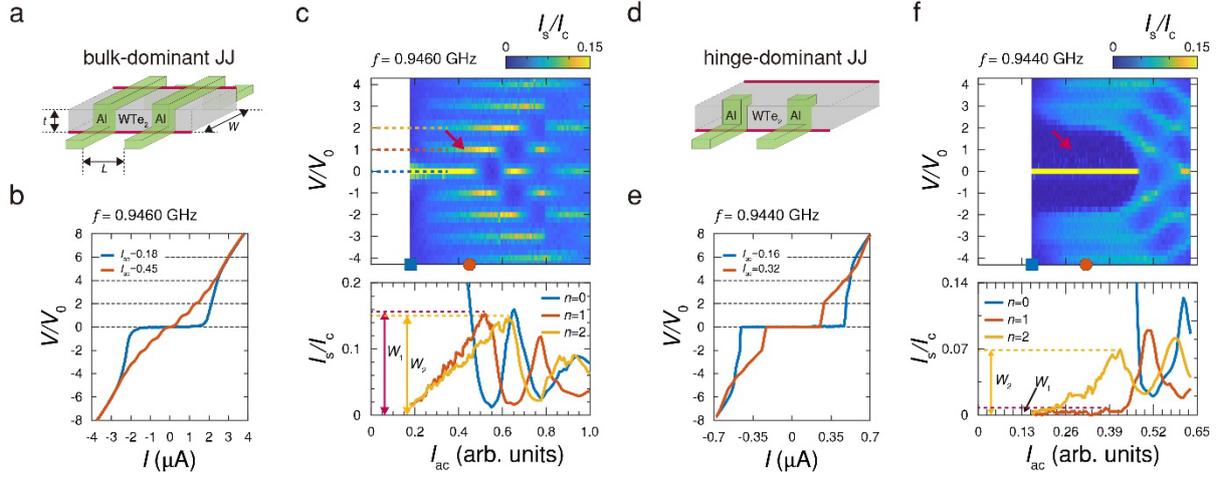

**Fig. 2 Shapiro steps in bJJ and hJJ. a, d,** Schematic representations of (**a**) a bJJ and (**d**) an hJJ. Topological hinge states (red solid lines) are connected to aluminium superconducting electrodes (green). $W$, $L$, and $t$ represent the junction width, junction length, and thickness of the JJ, respectively. **b, e,** Current ($I$)–voltage ($V$) characteristics of (**b**) the bJJ with a microwave frequency $f = 0.9460$ GHz and (**e**) the hJJ with $f = 0.9440$ GHz at different nominal microwave alternating current (ac) amplitudes ($I_{ac}$). $I$–$V$ measurements were performed by sweeping the bias current from negative to positive with constant current steps. The $I_{ac}$ values correspond to symbols in (**c**) and (**f**): square (blue) and circle (red). **c, f,** (Upper panels) Shapiro steps as a function of $I_{ac}$ and normalised voltage ($V/V_0$) for the bJJ (**c**) and hJJ (**f**). Here, $V_0 = hf/2e$, where $h$ denotes Planck's constant, and $e$ is the electron charge. The colour map represents the width of the current step ($I_s$) at a given normalised voltage step ($\Delta V/V_0 = 0.25$), normalised by the critical current ($I_c$). The normalised voltage $V/V_0 = 1$ is indicated by a red arrow. (Lower panels) Horizontal line cuts of (**c**) and (**f**) at $V/V_0 = 0$ (blue), $V/V_0 = 1$ (red), and $V/V_0 = 2$ (orange). $w_1$ and $w_2$ represent the maximum $I_s/I_c$ values for the first lobes of $V/V_0 = 1$ and $V/V_0 = 2$ steps, respectively.

The JJs were fabricated using a 13-nm-thick multilayer WTe$_2$ crystal, with the Josephson current flowing along the $a$-axis of the crystal. The junction widths of the bJJ and hJJ were 5.8 μm and 0.65 μm, respectively (Figs. 2a and d). Both these JJs had the same channel length of $L = 0.2$ μm. The critical current ($I_c$) and normal-state resistance ($R_N$) were 2.73 μA and 4.40 Ω for the bJJ and 0.53 μA and 25.7 Ω for the hJJ, respectively. Despite the differences in junction geometry between the bJJ and hJJ, the Josephson coupling strengths, represented by $I_c R_N$ products, were similar for both: 12.0 µV for the bJJ and 13.6 µV for the hJJ. Assuming ballistic conduction through the hinge states, the estimated superconducting coherence length $\xi = \hbar v_F / 2\Delta_{Al} \sim 0.84$ µm was considerably longer than $L$, indicating a short-junction limit. Here, we used the Fermi velocity of WTe$_2$[55] ($v_F = 3.1 \times 10^5$ m/s), along with the aluminium (Al) superconducting gap $(\Delta_{Al} = 1.763 \cdot k_B T_{c,Al} = 0.12$ meV$)$. This superconducting gap was calculated using the critical temperature of the aluminium electrode $(T_{c,Al} = 0.75$ K$)$ and Boltzmann's constant $(k_B)$. To confirm the presence of hinge states in the WTe$_2$ crystal, we analysed the magnetic field interference behaviour of $I_c$, as reported in our previous study[2] (see Fig. S1). Next, we performed Shapiro step measurements by recording the $I$–$V$ characteristics at 20 mK while irradiating the JJ devices with microwaves. The JJs exhibited small hysteresis in their $I$–$V$ characteristics: $I_c/I_r \sim 1.03$ for the bJJ and 1.14 for the hJJ. This indicates that the junctions are marginally underdamped, with estimated McCumber parameter $\beta = 2e I_c R^2 C / \hbar$ from RCSJ model of 1.1 for the bJJ and 1.6 for the hJJ (see Fig. S2). Here, $R$ and $C$ denote resistance and capacitance, respectively.

The measured Shapiro steps in the bJJ, which has a significant bulk contribution (Fig. 2a), resemble the ordinary Shapiro steps observed in conventional $2\pi$-periodic JJs. Figure 2b illustrates the $I$–$V$ characteristics of the bJJ under microwave irradiation at different nominal values of $I_{ac}$. The value of $I_{ac}$ was estimated from the microwave power $(P)$ at the microwave source output using the relation $I_{ac} \equiv 10^{[P/(20 \text{ dBm})]}$. As depicted in the figure, Shapiro steps gradually appear in the $I$–$V$ characteristics with increasing $I_{ac}$. In the upper panel of Fig. 2c, colours represent the step width $(I_s)$, normalised by $I_c$. All integer multiples of $V_0$ (*i.e.* voltage steps $V_n = n V_0$ for integer $n$) are highlighted in yellow. Notably, the first step $(V_{n=1})$, indicated by the red arrow, remains prominent across all frequencies (see Fig. S3), resembling the behaviour of ordinary Shapiro steps. The lower panel of Fig. 2c presents horizontal line cuts from the upper panel at $n = 0, 1,$ and $2$, marked by dashed lines. These cuts illustrate the evolution of the step width with $I_{ac}$. The amplitudes of the first and second steps evolve continuously, following Bessel function behaviour. Furthermore, the first lobe of the $n$ = 1 step peaks at a lower $I_{ac}$ than that of the $n = 2$ step. Moreover, the ratio of $w_1$ to $w_2$ $(Q_{12} = w_1/w_2)$ is 1.04, which is greater than one. Here, $w_1$ and $w_2$ denote the amplitude maxima of the first lobes of the voltage steps $n = 1$ and $n = 2$, respectively. Overall, in the bJJ, transport through topologically trivial bulk states may overshadow contributions from hinge states, leading to ordinary Shapiro steps.

In the hJJ (Fig. 2d), which has a larger hinge-to-bulk contribution ratio than the bJJ, the first Shapiro step is entirely absent as shown in Fig. 2e. Similar results were observed from two additional devices (see Fig. S4). In the upper panel of Fig. 2f, the region near $V/V_0 = 1$, marked by the red arrow, appears dark blue, suggesting that $I_s$ is close to zero. The lower panel of Fig. 2f illustrates that the $I_s$ of the first step remains nearly zero until $I_{ac}$ approaches 0.43, where the second lobe of the $n = 1$ step begins to emerge. Conversely, the $I_s$ of the second step evolves continuously from a lower $I_{ac}$. Quantitative analysis reveals that $w_1 \sim 0.004$ and $w_2 \sim 0.07$, yielding $Q_{12} = 0.057$, a value considerably smaller than one. In this measurement, $w_1 \times I_c \sim 2$ nA is comparable to the minimum current step (2 nA).

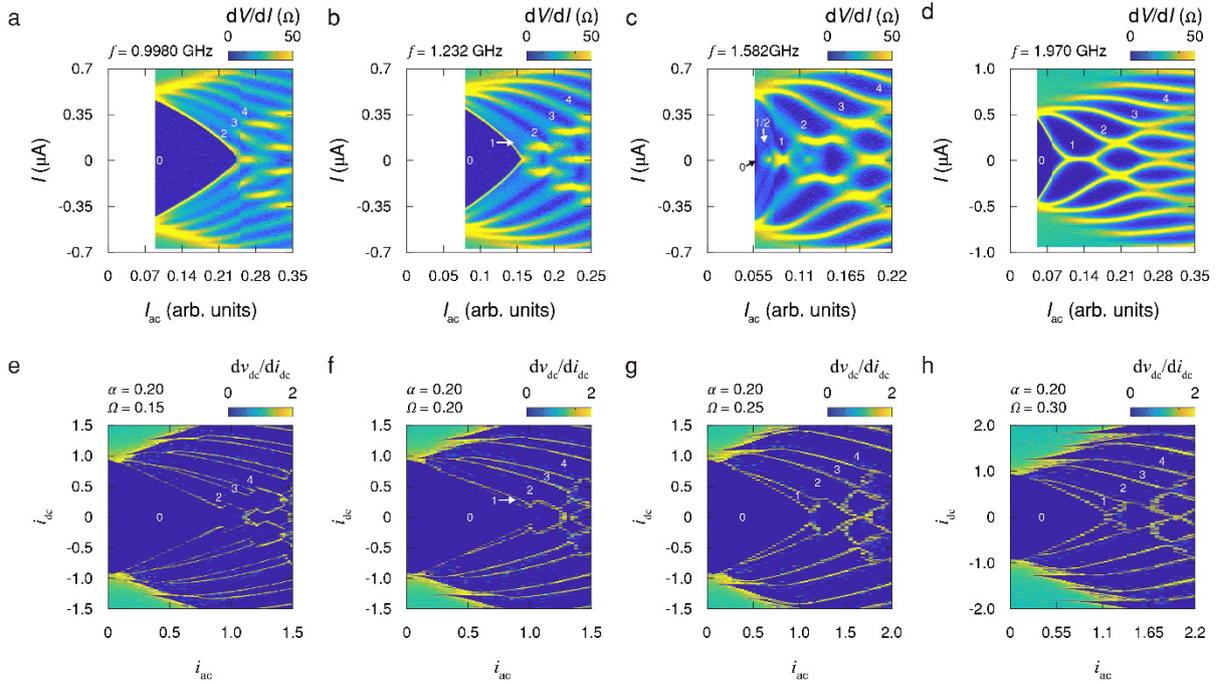

**Fig. 3 Frequency dependence of Shapiro steps in the hJJ. a–d** Shapiro steps measured in the hJJ as a function of $I_{ac}$ and bias current $I$ at different microwave frequencies: (**a**) $f = 0.998$ GHz, (**b**) $f = 1.232$ GHz, (**c**) $f = 1.582$ GHz, and (**d**) $f = 1.970$ GHz. Colours indicate the differential resistance. **e, f,** Numerically calculated Shapiro steps for $\beta = 1.6$ and $\alpha = 0.20$ as functions of $i_{ac}$ and bias current ($i_{dc}$) at different dimensionless frequencies ($\Omega = hf/2eI_cR$): (**e**) $\Omega = 0.15$, (**b**) $\Omega = 0.20$, (**c**) $\Omega = 0.25$, and (**d**) $\Omega = 0.30$. Here, $\alpha$ denotes the ratio of the 4π-periodic Josephson current to the 2π-periodic Josephson current, and $\beta = 2eI_cR^2C/\hbar$ represents the McCumber parameter. $v_{dc}$ denotes the dc voltage applied across the junction normalised by $I_cR$, and $i_{ac}$ and $i_{dc}$ represent the normalised ac and dc current,

respectively.

We investigated the dependence of Shapiro steps on microwave frequency in the hJJ and compared the results with the outcomes of numerical calculations. Figure 3 presents experimentally measured Shapiro steps alongside those derived from numerical calculations based on the RCSJ model, using the CPR of $I_J(\phi) = I_{4\pi} \sin(\phi/2) + I_{2\pi} \sin(\phi)$. Here, $I_J$ represents the Josephson current, and $I_{2\pi}$ and $I_{4\pi}$ denote the $2\pi$- and $4\pi$-periodic Josephson currents, respectively. The numerical calculations were performed using $\beta = 1.6$, estimated from the I–V characteristics of the hJJ, and $\alpha = 0.20$, which best matches the experimental data (see Fig. S5 for numerical results at different $\alpha$). Here, $\alpha = I_{4\pi}/I_{2\pi}$ represents the ratio of the $4\pi$-periodic Josephson current to the $2\pi$-periodic one. In Fig. 3a, the first lobe of the $n = 1$ step is entirely absent. As $f$ increases, the first Shapiro step becomes faintly visible. At $f = 1.232$ GHz (Fig. 3b), the first Shapiro step, indicated by the white arrow, faintly appears near the $n = 0$ step. As $f$ increases further, the Shapiro steps evolve into a standard form, displaying all integer steps. This behaviour is consistent with theoretical predictions[43-45] (Figs. 3e–h) and experimental findings from previous studies[47, 48, 50-54] (see Supplementary Fig. S6 for Shapiro steps as a function of $I_{ac}$ and $V/V_0$).

The experimental results show broader boundaries between adjacent steps compared to the numerical calculations. This discrepancy is likely attributed to Joule heating, arising from both the dc and microwave radiation introduced through the antenna. The resulting increase in electron temperature alters the phase dynamics of particles in the JJs. Consequently, the slopes of the plateaux where Shapiro steps occur become less distinct, and the sharp voltage transitions between steps appear smoother. To mitigate the effects of Joule heating, we focus on the ratio between the first lobes of the $n = 1$ and $n = 2$ steps, represented by $Q_{12}$, which is evaluated under conditions of low dc and ac values. Notably, this low-current regime underscores the significant influence of the $4\pi$-periodic CPR on Shapiro steps, as supported by previous studies[46].

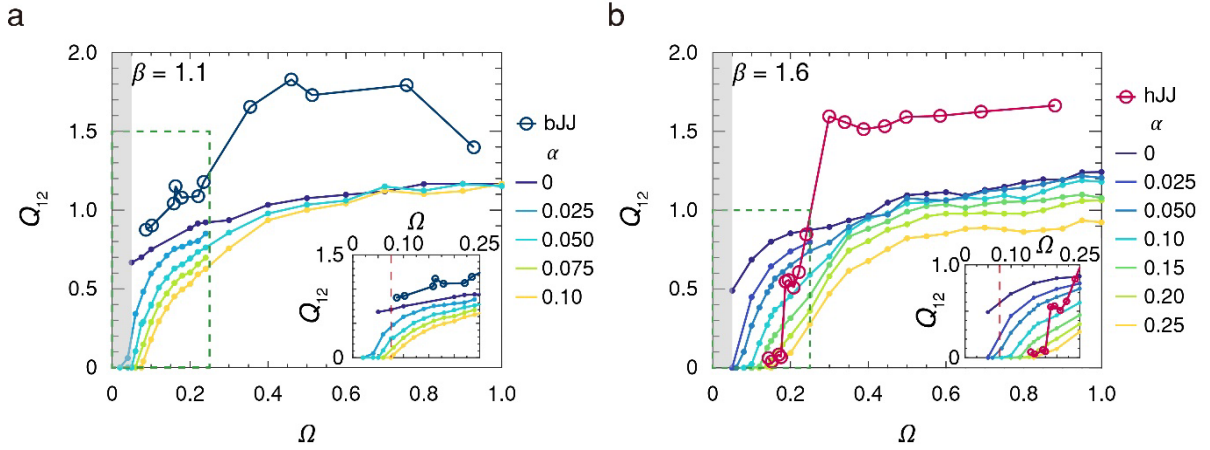

**Fig. 4. Microwave frequency dependence of $Q_{12}$.** Microwave frequency dependence of $Q_{12}$ as a function of the dimensionless frequency for (**a**) $\beta = 1.1$ and (**b**) $\beta = 1.6$. Here, $\alpha$ denotes the ratio of the 4π-periodic Josephson current to the 2π-periodic current, $\beta = 2eI_cR^2C/\hbar$ represents the McCumber parameter, $Q_{12} = w_1/w_2$, and $\Omega = hf/2eI_cR$ represents the dimensionless frequency. The solid lines represent the numerically calculated $Q_{12}$, while the solid lines with circles indicate experimental results. The grey area marks the frequency range that cannot be measured experimentally. The green dashed box highlights the zoomed-in region shown in the inset. Inset: Enlarged view of the region from $\Omega = 0$ to $\Omega = 0.25$. The red dashed line in the inset corresponds to $\Omega = 0.075$.

Figures 4a and 4b display the experimentally measured $Q_{12}$ values as functions of $\Omega$ for the bJJ and hJJ, respectively, alongside the numerically calculated values at various $\alpha$. Both experimental and numerical results extend down to $\Omega \sim 0.05$ (corresponding to $f \sim 0.3$ GHz), below which the Shapiro step voltage $V_0 < 0.6$ μV becomes too small to be measured experimentally. As $\Omega$ decreases, $Q_{12}$ approaches zero for $\alpha \neq 0$ or approaches to a non-zero value of approximately 0.7 for $\beta = 1.1$ and 0.5 at $\beta = 1.6$ when $\alpha = 0$. Moreover, $Q_{12}$ tends to decrease as $\alpha$ or $\beta$ increases for a given $\Omega$ (see Supplementary Fig. S7). At higher frequencies ($\Omega > 0.3$), the experimentally measured $Q_{12}$ is consistently larger than the numerical values, likely owing to Joule heating, in both the bJJ and hJJ. The Joule heating power ($P$) is proportional to $n^2$ ($P \propto V_n^2 \propto n^2 f^2$) at the $n$-th Shapiro step; hence, $w_2$, measured at the $n = 2$ step, is more affected by elevated electron temperature than $w_1$, measured at $n = 1$. This results in an overestimation of $Q_{12}$, which would otherwise diminish at lower $f$. The value of $\Omega$ below which $Q_{12}$ vanishes is used as a proxy for estimating $\alpha$, following the approach of previous studies[46, 50]. For the hJJ, this yields $\alpha_{hJJ} = 0.2$, which corresponds to $I_{4\pi} = 93$ nA. This value is consistent with that ($I_{4\pi}$= 95 nA) derived from the magnetic field interference behaviour observed in the same device (see Fig. S1e). Similar $\alpha = 0.10-0.25$ values are obtained for other hJJs (see Fig. S8). The Josephson current mediated by a single hinge state is estimated as $I_{J,h} = 15$ nA, considering the short ballistic junction limit $I_{J,h}R_{N,h} = \pi\Delta_{Al}/e$, where $R_{N,h} = h/e^2$ denotes the normal-state resistance for a single hinge state. The $I_{4\pi}$ value derived from the Shapiro step analysis likely originates from multiple hinge states located at the terraces of the WTe$_2$ crystal, which may have formed during mechanical exfoliation. Assuming that each of the two crystal edges (upper and lower) of the bJJ has an $I_{4\pi}$ comparable to that of the hJJ, $\alpha_{bJJ} = 0.07$ can be estimated for the bJJ. This indicates that the complete missing of the first Shapiro step ($Q_{12} = 0$) may also occur in the bJJ at $f < 0.4$ GHz. However, at such low microwave frequencies, $V_0$ becomes as small as 0.8 µV, making experimental measurement extremely complex. Consequently, potential topological hinge states in the bJJ are obscured by dominant contributions from trivial conduction states.

Our next analysis focused on determining whether the complete missing of the first Shapiro step in the hJJ originates from topologically trivial mechanisms. The missing of the first Shapiro step can result from topologically trivial factors such as Landau–Zener transitions (LZT) or the underdamped nature of JJs. First, the LZT is a non-adiabatic transition triggered by microwaves, which can create an ABS bandgap near $\phi \sim \pi$. The probability of the LZT[56, 57] is defined as $P_{LZT} = \exp\left(-\pi\frac{(1-D)\Delta}{eV_{LZT}}\right)$, where $D$ denotes the junction transparency, and $V_{LZT}$ represents the junction voltage at $\phi = \pi$. Given the similar Josephson coupling strengths of the bJJ and hJJ, $D$ is also likely comparable[58]. Moreover, Shapiro steps for both junctions were measured within a similar microwave frequency range. Therefore, if the LZT were responsible for the first Shapiro step missing, this phenomenon should have been observed in both devices. Second, we evaluated the underdamped nature of the JJs using our previous RCSJ model calculations[46]. At sufficiently low $\Omega$, $Q_{12}$ decreases with increasing $\beta$, even when $\alpha = 0$. However, in the absence of a 4π-periodic Josephson current, the complete missing of the first Shapiro step ($Q_{12} = 0$) does not occur for $\beta \leq 3$. Given that $\beta = 1.6$ for the hJJ is considerably smaller than three, the first Shapiro step missing observed in the hJJ cannot be explained by the underdamped nature of the junction.

In this study, we fabricated two types of WTe$_2$ JJs: a bJJ and an hJJ, with current flowing along the *a*-axis of the crystal. To investigate the topological nature of the hinge states, we analysed the Shapiro step behaviour of these junctions. Under microwave irradiation, the hJJ, with reduced conduction contributions from topologically trivial surface and bulk states, exhibits the fractional Josephson effect, characterised by the first Shapiro step missing. In contrast, the bJJ does not exhibit the fractional Josephson effect, which is attributed to the significant conduction contributions from topologically trivial states. We further examined the microwave frequency and power dependencies of the Shapiro steps and provided quantitative explanations using numerical calculations based on the RCSJ model. Notably, the estimated value of the 4π-periodic Josephson current in the hJJ closely matches the Josephson current along the hinges, as determined from the magnetic field interference pattern. This result further supports the topological nature of the hinge states in $WTe_2$. While our findings suggest the topological nature of the hinge states, they remain inconclusive and require further investigation, such as Josephson radiation[59-61] or bimodal switching current distribution[62].

## Methods

**Device fabrication.** A multilayer $WTe_2$ crystal was mechanically exfoliated onto a silicon substrate with a 280-nm-thick insulating layer of $SiO_2$. The crystal thickness was confirmed to be 13 nm using atomic force microscopy (using an XE-7 from Park Systems). To remove any contamination from the topmost layers of $WTe_2$ flakes, *in situ* argon ion etching was conducted before metal deposition to eliminate $WTe_2$ surface layers potentially oxidised in ambient air. Ti/Al/Au (5 nm/60 nm/5 nm) electrodes were deposited through evaporation to facilitate the study of superconducting transport. The Ti layer improved the adhesion of the electrodes to $WTe_2$, while the Au layer protected the aluminium superconducting electrode from oxidation. During evaporation, the chamber pressure was maintained below $2 \times 10^{-7}$ Torr. To minimise degradation, the multilayer $WTe_2$ crystal was exposed to poly(methyl methacrylate) polymer only once during the fabrication process.

## Acknowledgement

Half of this research was supported by the Nano and Material Technology Development Program through the National Research Foundation of Korea (NRF), funded by the Ministry of Science and ICT (No. RS-2024-00444725). Additional support was provided by other NRF

grants (Nos. RS-2022-NR068223, RS-2024-00393599, RS-2024-00442710) and the ITRC program (IITP-2025-RS-2022-00164799), also funded by the Ministry of Science and ICT. Further funding was received from the Samsung Science and Technology Foundation (SSTF-BA2101-06, SSTF-BA2401-03) and Samsung Electronics Co., Ltd. (IO201207-07801-01). Y.-B. Choi acknowledges support from the POSTECHIAN fellowship and the POSTECH Alchemist program.

# Supplementary Information for

# Unveiling Topological Hinge States in the Higher-Order Topological Insulator WTe$_2$ Based on the Fractional Josephson Effect

Yong-Bin Choi[1,*], Jinho Park[1,2,*], Woochan Jung[1], Sein Park[1], Mazhar N. Ali[3,4], and Gil-Ho Lee[1,5,†]

[1]Department of Physics, POSTECH, Pohang, Republic Korea

[2]Department of Mechanical Engineering, Columbia University, New York, NY 10027, USA

[3]Kavli Institute of Nanoscience, Delft University of Technology, Delft, Netherlands

[4]Department of Quantum Nanoscience, Delft University of Technology, Delft, Netherlands

[5]Asia Pacific Center for Theoretical Physics, Pohang, Republic of Korea

[*]These authors contributed equally.

[†]Corresponding authors: lghman@postech.ac.kr (G.-H.L.)

# S1. Magnetic field interference of bJJ and hJJ

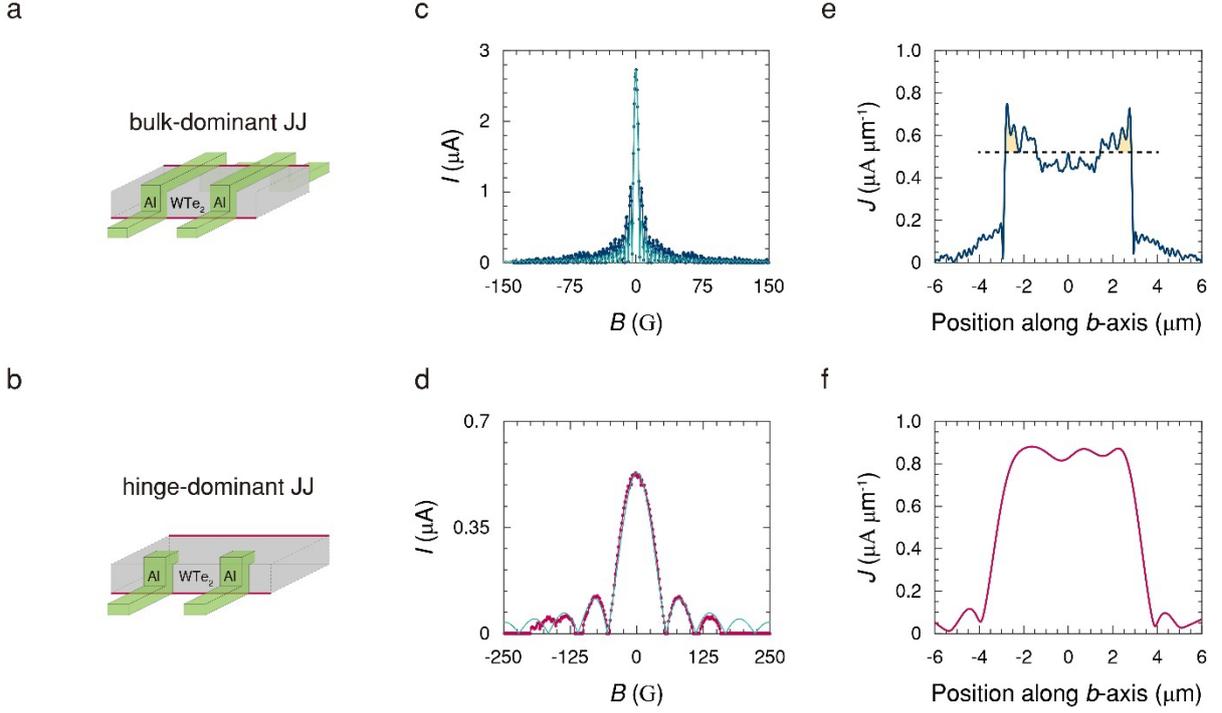

**Fig. S1. Magnetic interference pattern. a, b,** Schematics of devices for (**a**) bulk-dominant Josephson junction (bJJ) and (**b**) hinge-dominant JJ (hJJ). **c, d,** Magnetic interference patterns for bJJ (**c**) and hJJ (**d**) as a function of differential resistance. The green lines represent a single-slit Fraunhofer pattern. **e, f,** Extracted spatial distributions of the Josephson current density $J$ for bJJ (**e**) and hJJ (**f**). The yellow area represent enhanced current density, with the baseline consisting of the bulk contribution (black dashed line)

The presence of hinge states in the multilayer $WTe_2$ crystal is demonstrated using Josephson junctions (JJs) with Al superconducting electrode, where the Josephson current flow along the $a$-axis of the crystal (Figs. S1a and b). A non-vanishing magnetic field interference, previously reported[1-4], is observed. As shown in Fig. S1c, the magnetic field interference pattern, known as Fraunhofer pattern (FP), persists up to $\pm 150$ G for the bulk-dominant JJ (bJJ). The period of the critical current ($I_c$) oscillation, $\Delta B=4.5$ G, corresponds well with the flux quantum threading an effective junction area $WL_{\text{eff}}$, where $L_{\text{eff}} = L + 2L'$, $L \simeq 200$ nm, and $L'$ represents half the width of the Al electrodes.

The critical current of Al superconducting electrodes ($I_c^{\text{Al}}$) are influenced by the critical magnetic field $B_c^{\text{Al}}$, following the relation $I_c^{\text{Al}}/I_{c0}^{\text{Al}} \simeq \sqrt{1 - (B/B_c^{\text{Al}})^2}$, where $I_{c0}^{\text{Al}}$ represents critical current at zero temperature. This relation illustrates how the critical current decreases as the applied magnetic field $B$ approaches the critical field $B_c$. A similar dependence holds for the JJs, described by $I_c/I_{c0} \simeq \sqrt{1-(B/B_c)^2}$. As the magnetic field increases, the Josephson coupling weakens, and the non-vanishing behaviour gradually diminishes. Eventually, the non-vanishing behaviour follows ordinary behaviour at higher magnetic field. By applying an inverse Fourier transform to the FP, the spatial current distribution is obtained (Fig. S1e). This edge-enhanced spatial current distribution represents the presence of hinge states. From the spatial distribution, the enhanced Josephson current along a single edge of the junction ($I_{J,side}$) is determined to be 95±27 nA for the left edge and 88±28 nA for the right edge. Here, the bulk current level, represented by the dashed line, corresponds to the average value between the left and right edge peaks. For the hinge-dominant JJ (hJJ), the FP shows ordinary behaviour, with an $I_c$ oscillation period of $\Delta B$=52 G. The inverse Fourier transform results a uniform current density (Fig. S1f), which can be interpreted as arising from a single hinge state. However, it is challenging to determine whether this ordinary behaviour originates from a single hinge state or unconventional geometrical effects.

## S2. Resistively and capacitively shunted junction (RCSJ) model

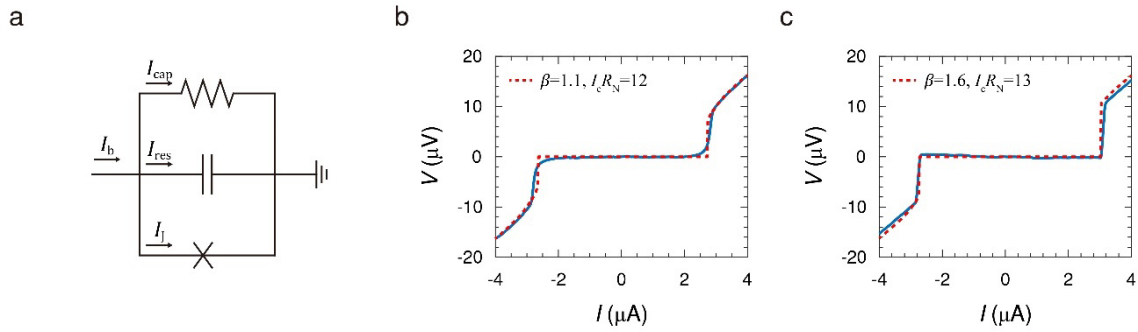

**Fig. S2. Current-voltage characteristics. a,** Schematic diagram of the resistively and capacitively shunted junction (RCSJ) model, illustrating a lumped circuit elements, which consist of a resistor and a capacitor in parallel with the Josephson element. Here, $I_b$, $I_{res}$, $I_{cap}$, and $I_J$ represent bias current and the current flow through the resistor, capacitor and Josephson junction, respectively. **b, c,** Current-voltage characteristics of **(a)** bJJ and **(b)** hJJ, respectively. The red dashed lines represent fitting curves using the RCSJ model.

The RCSJ model is an electrical circuit model used to describe Josephson dynamics. The JJ can be represented using lumped circuit elements, where Josephson element is shunted by a resistive and a capacitive element in parallel (Fig. S2a). According to Kirchhoff's circuit laws, the bias current can be expressed as $I_b = I_{res} + I_{cap} + I_J$. Using the Josephson relations $I_J = I_c \sin\phi$ and $V_J = \frac{\hbar}{2e}\frac{d\phi}{dt}$, the equation of motion can be formulated as a second-order differential equation in terms of $\phi$,

$$I_b = \frac{\hbar C}{2e}\frac{d^2\phi}{dt^2} + \frac{\hbar}{2eR}\frac{d\phi}{dt} + I_c \sin\phi. \tag{S1}$$

Here, $\phi$ represents macroscopic quantum phase difference between two superconductors, $\hbar$ is Planck's constant $h$ divided by $2\pi$, $e$ is the electron charge, and $C$ and $R$ denote the capacitance and resistance of the circuit, respectively. Equation (1) can be simplified into dimensionless form[5-7] by introducing the dimensionless time variable $\tau$, where $t = \tau\omega_p^{-1}$, and $\omega_p$ is plasma frequency of the junction given by $\omega_p = \sqrt{2eI_c/\hbar C}$. The resulting dimensionless equation is

$$\frac{d^2\phi}{d\tau^2} + \frac{1}{\beta}\frac{d\phi}{d\tau} + \sin\phi - \frac{I_b}{I_c} = 0, \tag{S2}$$

where $\beta$ is McCumber parameter, defined as $\beta = \frac{2eI_c R^2 C}{\hbar}$. Equation (S2) is analogous to the equation of motion of a damped harmonic oscillator, where the second term represents the damping force, and the sine term corresponds to the restoring force. Here, the McCumber parameter $\beta$ determines the damping rate. When $\beta < 1$, the junction is overdamped, and the current-voltage (I-V) characteristics exhibit non-hysteretic behaviour. Conversely, when $\beta > 1$, the junction is underdamped, leading to hysteretic behaviour in the I-V characteristics. The I-V characteristics, obtained by solving equation (S2), are compared with experimental measurements to estimate $\beta$ for both the hJJ and the bJJ, as shown by the red dashed lines in Figs. S2b and c.

# S3. Frequency dependence of the Shapiro steps in bJJ

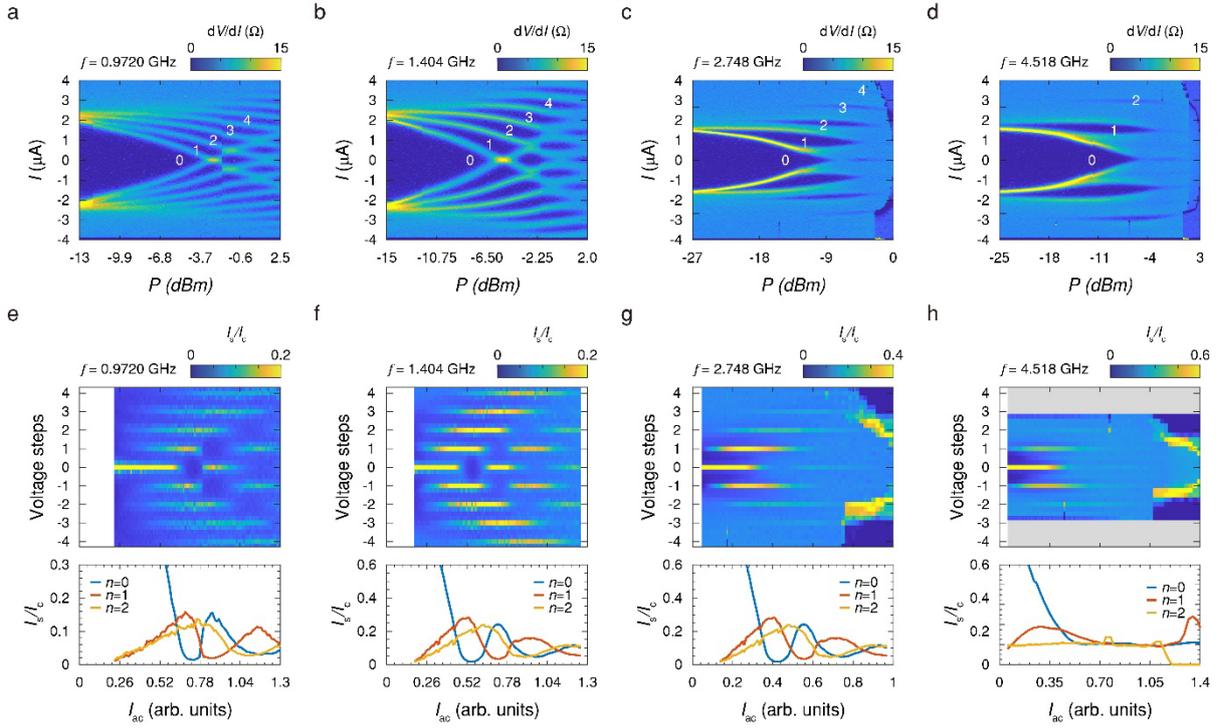

**Fig. S3. Frequency dependencies of Shapiro steps in bulk-dominant Josephson junction. a-d,** Shapiro steps as a function of microwave power ($P$) and bias current ($I$) at different microwave frequencies $f$. Colour represents the differential resistance. **e-h,** (Upper panels) Shapiro steps as a function of $I_{ac}$ and voltage steps normalized by normalized $V_0 = hf/2e$, where $h$ is the Planck's constant and $e$ is the electron charge. Colour represents the width of current step ($I_s$) normalized by the critical current $I_c$. (Lower panels) Horizontal linecut from the middle panels at $V/V_0 = 0$ (blue), $V/V_0 = 1$ (red), and $V/V_0 = 2$ (orange).

As shown in Fig. S3, the first Shapiro step missing is not observed in the bJJ, regardless of the microwave frequency $f$. As shown in the upper panels and lower panels of Fig. S3e-h, all integer steps are distinctly highlighted, and $w_1$ is higher than $w_2$. As $f$ increase, higher voltage steps disappear due to the Joule heating. For instance, at $f$=2.748 GHz, the $n = 2$ step becomes less distinct compared to the $n = 1$ step. Moreover, at $I_{ac} \sim 0.7$, it can be observed that Josephson coupling is not formed due to Joule heating.

## S4. Shapiro steps in additional devices

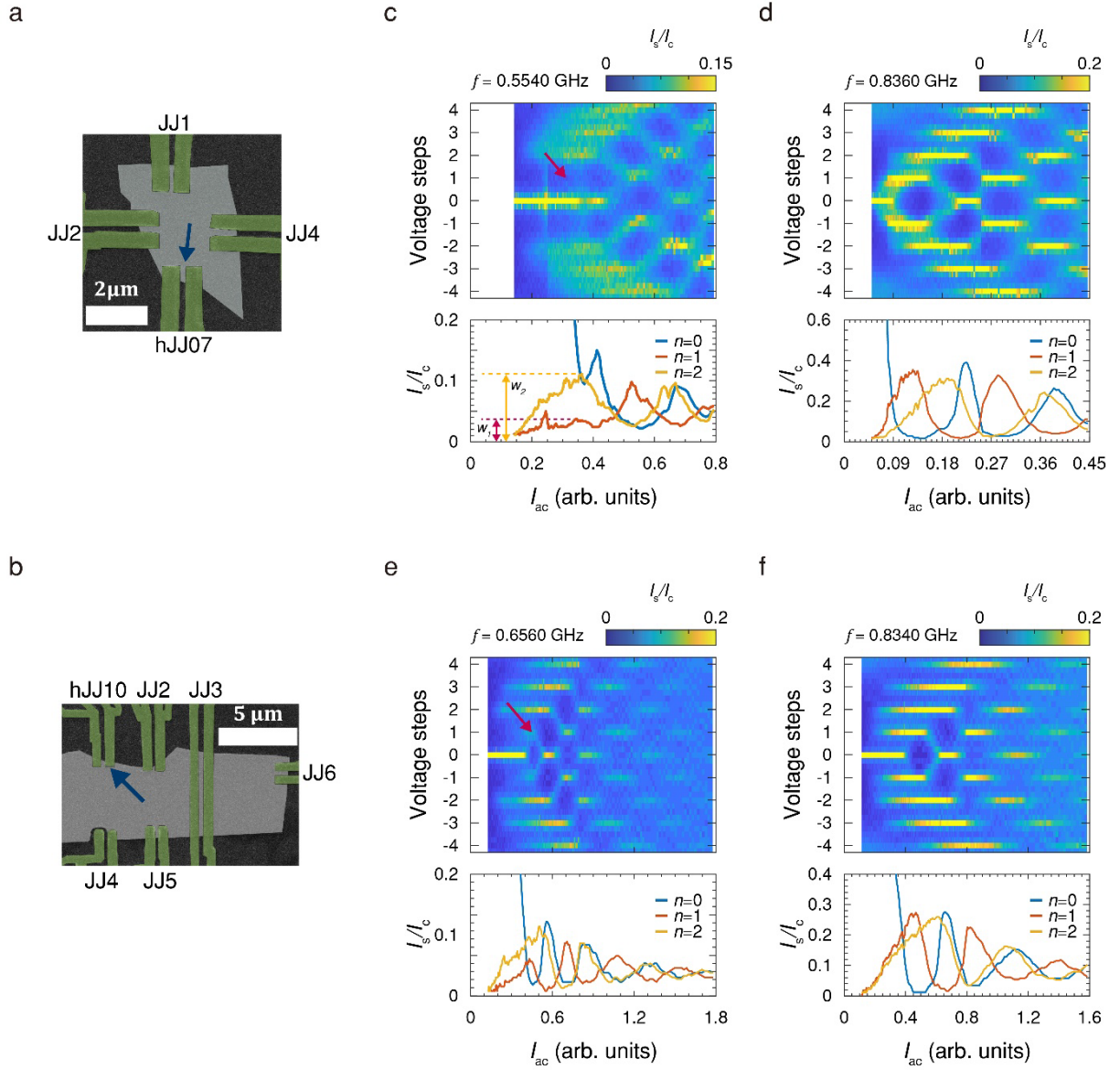

**Fig. S4. More device of THS 07 and THS 10. a, b,** False-colour scanning electron micrograph of THS 07 and 10, respectively. The blue arrow indicates the measured junction in each device. **c-f,** (upper panels) Shapiro steps as a function of $I_{ac}$ and voltage steps normalized by $V_0 = hf/2e$, measured in devices THS 07 (**c, d**) and THS 10 (**e, f**). (lower panels) Horizontal linecut of **c, d, e,** and **f** respectively, at $V/V_0 = 0$ (blue), $V/V_0 = 1$ (red), and $V/V_0 = 2$ (orange). $w_1$ and $w_2$ denote the maximum $I_s/I_c$ values of the first lobe of the $V/V_0 = 1$ and $V/V_0 = 2$ steps, respectively.

As shown in Fig. S4, Shapiro step measurements were conducted on two separate devices, named THS 07 and THS 10. These devices were fabricated using multilayer WTe$_2$ crystal, thickness of 10 nm and 17 nm for THS 07 and THS 10, respectively. We focused on the junction where the Josephson current flows along the *a*-axis of the crystal, as indicated by the blue arrow in Fig. S4a and S4b. These junctions are labelled hJJ07 and hJJ10, respectively. The junction widths of hJJ07 and hJJ10 are 0.47 µm and 0.35 µm, respectively, and both JJs have same channel length $L = 0.2$ µm. The $I_c$ and the normal-state resistance ($R_N$) of hJJ07 are 0.089 µA and 61.5 Ω, respectively, while for hJJ10, $I_c$ and $R_N$ are 0.34 µA and 18.9 Ω, respectively, resulting in $I_c R_N = 5.47$ µV for hJJ07 and $I_c R_N = 6.43$ µV for hJJ10..

The Shapiro step measurement was conducted by measuring the *I-V* characteristics at 20 mK while microwave is irradiated onto the JJ devices. As shown in the Fig. S4, for the hJJ07, the first Shapiro step, as indicated by the red arrow, is absent, while higher steps are clearly visible and highlighted in yellow. In lower panel of Fig. S4c, $w_2$ is much larger than $w_1$. When $f$ is increased, the first step starts to appear. At $f$=0.8360 GHz, as shown in Fig. S4d, ordinary Shapiro step is observed, and $Q_{12}$ becomes larger than 1, indicating that the $4\pi$-periodic current is dominated by $2\pi$-periodic current. For the hJJ10, the suppression of the first Shapiro step is observed. As shown in the Fig. S4e, the first step slightly remains and $w_1$ is considerably large, but $Q_{12} = \frac{0.091}{0.17} = 0.54$ is far below 1. Similar to the other hJJ, ordinary Shapiro steps are observed when $f$ is increased, as shown in Fig. S4f. This microwave frequency dependence well corresponds to the RCSJ model calculations.

## S5. Numerical calculation of the Shapiro steps

When a Josephson junction is irradiated with microwave, the bias current applied across the junction consists of both dc and ac components, expressed as $I_b = I_{dc} + I_{ac} \cos(\omega_{ac} t)$. The equation (S2) can be rewritten as equation (S3)

$$\beta \ddot{\phi} + \dot{\phi} + i_J = i_{dc} + i_{ac} \cos(\Omega \tau), \qquad (S3)$$

with dimensionless parameters $i_x = I_x / I_c$ and $\Omega = \frac{\hbar \omega_{ac}}{2 e I_c R}$. Here, $\dot{\phi}$ and $\ddot{\phi}$ represent $\frac{d\phi}{d\tau}$ and $\frac{d^2\phi}{d\tau^2}$, respectively, with dimension less time with $\tau = \frac{2 e I_c R}{\hbar} t$. Note that we chose $\tau = \frac{2 e I_c R}{\hbar} t$ instead of $\tau = \omega_p t$ as is in equation (S2) to focus on the Shapiro steps, where the normalized Shapiro step $V_0 / I_c R = n\Omega$ is explicit.

To explore the Shapiro steps in the presence of a $4\pi$-periodic Josephson current, we assume

$i_J = i_{4\pi}\sin(\phi/2) + i_{2\pi}\sin(\phi)$ where $\alpha$ represents the ratio of 4π-periodic to 2π-periodic Josephson current, $\alpha = I_{4\pi}/I_{2\pi}$. By solving equation (S3) numerically for fixed values of $\alpha$ and $\Omega$, and varying $i_{dc}$ and $i_{ac}$, we obtain a differential resistance ($dv_{dc}/di_{dc}$) map that illustrates how the Shapiro steps evolve with the bias currents, where $v_{dc}$ represents dc voltage applied across the junction normalized by $I_cR$. Variations in $\alpha$ and $\Omega$ are shown in Fig. S5 and Fig. 3. The numerical calculations were conducted at a fixed value of $\beta = 1.6$, which was estimated from the current-voltage characteristics measured in the hJJ.

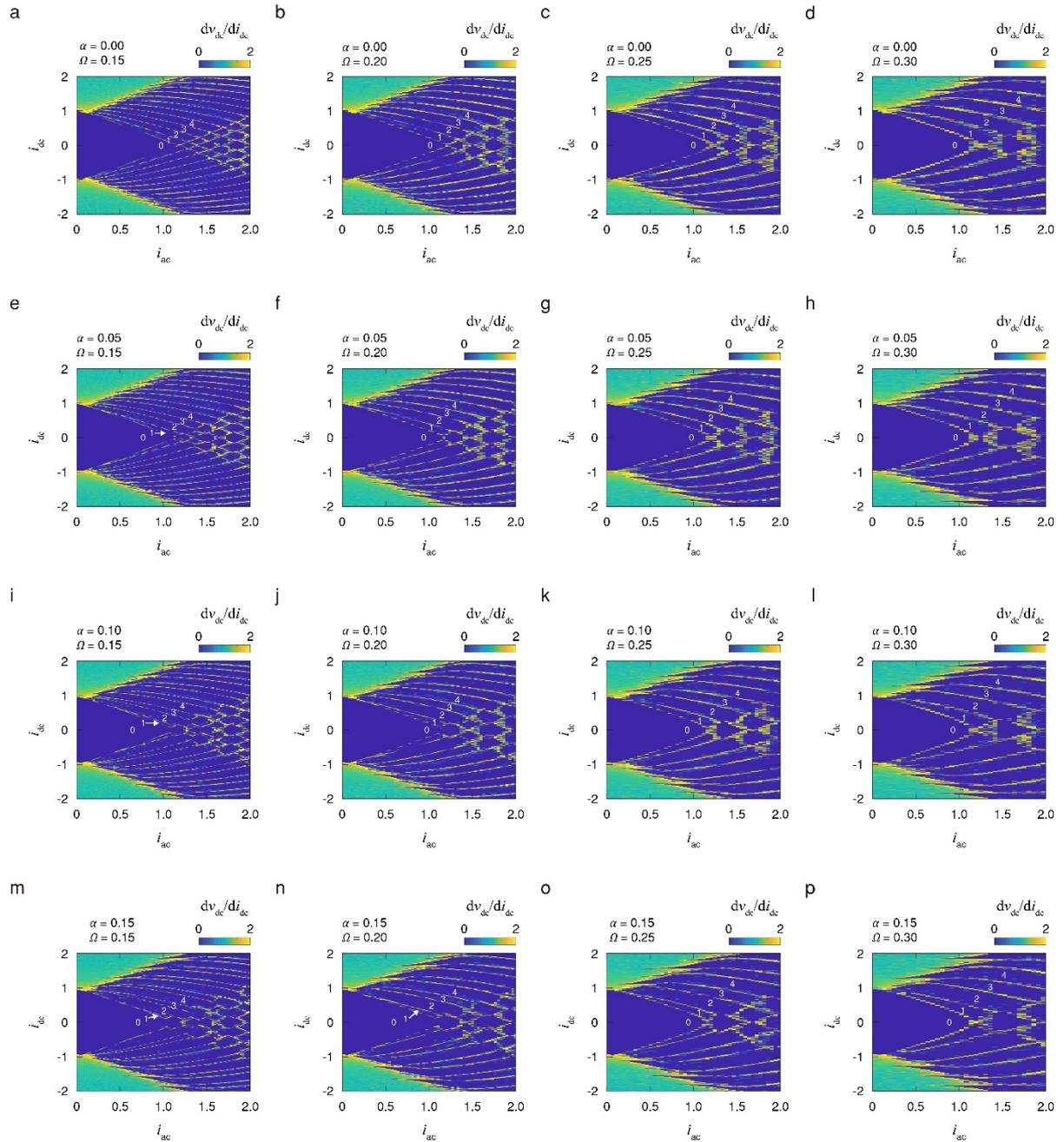

**Fig. S5. Numerically calculated Shapiro steps varying with $\alpha$ and $\Omega$.** Numerically

calculated Shapiro steps for $\beta=1.6$ as a function of $i_{ac}$ and bias current $i_{dc}$ with varying $\alpha$ and dimensionless frequency $\Omega = hf/2eI_cR$. Here, $\alpha$ represents the ratio of the $4\pi$-periodic Josephson current to $2\pi$-periodic one, $v_{dc}$ represents dc voltage applied across the junction normalized by $I_cR$, $i_{ac}$ and $i_{dc}$ are denoted normalized ac and dc current, respectively. Numbers represent $n$-th voltage steps.

## S6. Frequency dependence of the Shapiro steps in hJJ

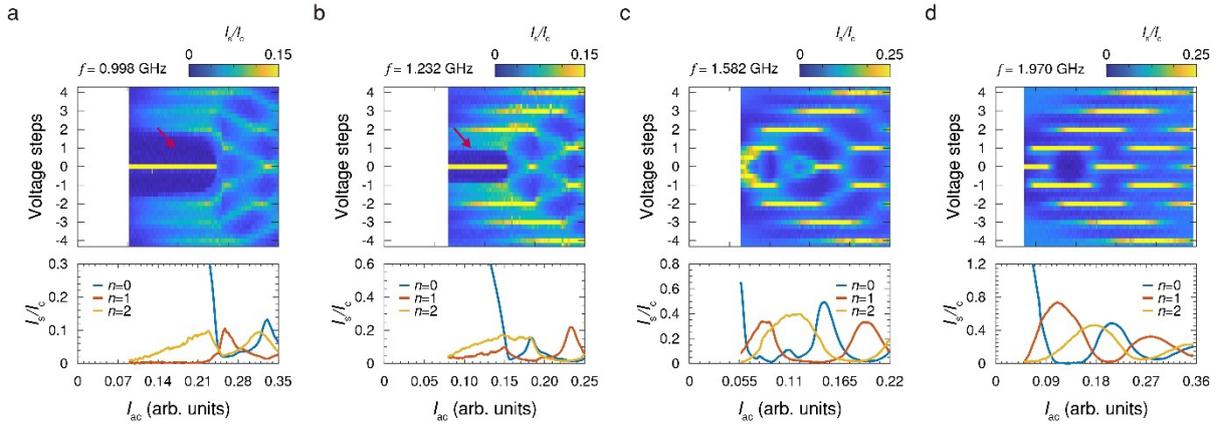

**Fig. S6. Frequency dependence of Shapiro steps in a hinge-dominant Josephson junction (hJJ).** (Upper panels) Shapiro steps measured in a hJJ as a function of $I_{ac}$ and normalized voltage $V/V_0$ at different microwave frequencies $f$, (**a**) $f$=0.998 GHz, (**b**) $f$=1.232 GHz, (**c**) $f$=1.582 GHz, and (**d**) $f$=1.970 GHz. (Lower panels) Horizontal linecut corresponding to $n = V/V_0 = 0, 1, 2$ from the upper panel.

## S7. Dependence of numerically estimated $Q_{12}$ on $\alpha$, $\beta$, and $\Omega$

The solution of equation (S3) can be understood qualitatively as the phase particle traversing an oscillating washboard potential, with damping proportional to its phase velocity $\dot{\phi}$. The mean slope of the washboard potential is given by $i_{dc}/\beta$, while the oscillation of the washboard is determined by the amplitude $i_{ac}/\beta$ and the dimensionless frequency $\Omega$ introduced by the microwave irradiation. The average phase velocity $\langle\dot{\phi}\rangle$, which corresponds to the dc voltage measured in the experiment, is proportional to the number of wells in the washboard potential that the phase particle traverses per a cycle of the oscillation of the washboard. Despite some changes in the mean slope of the washboard potential, the number of wells traversed per cycle can remain constant, giving the quantized voltage steps in the I-V curves called Shapiro steps.

For given $i_{dc}$ and $i_{ac}$, the average phase velocity, $\langle\dot{\phi}\rangle$, can vary depending on the junction parameters, $\alpha$, $\beta$ and $\Omega$. To illustrate this, we numerically calculated phase particle dynamics under microwave irradiation for various $\alpha$, as shown in Fig. S7 and Video S1. The parameter $\alpha$ modulates the washboard potential by adding $4\pi$-periodic contribution, making the washboard potential wells deep and shallow alternately (Fig. S7a). Figure S7b (c) depicts the Shapiro map for $\beta = 1.6$, $\Omega = 0.15$, and $\alpha = 0$ ($\alpha = 0.4$). For a representative example, we focus on the condition of $i_{ac} = 0.93$, $i_{dc} = 0.3$ where $n = 1$ step emerges for $\alpha = 0$ but $n = 2$ step emerges for $\alpha = 0.4$. Figure S7d shows the ac current applied across the junction as a function of time. The phase particle traversal and the oscillating washboard potential at specific time points marked in Fig. S7d are captured in Fig. S7e-j. For $\alpha = 0.4$, the phase particle passes the shallow well easily and then be trapped in the next deep well, so it traverses two wells per cycle. In contrast, for $\alpha = 0$, the phase particle traverses only one well per cycle. As $\alpha$ increases, the disparity between shallow and deep wells in the washboard potential become pronounced. This enhances the likelihood of the phase particle passing through the shallow well and getting trapped in the deep well, leading to the dominance of even-numbered Shapiro steps. Consequently, $Q_{12}$ decreases with increasing $\alpha$.

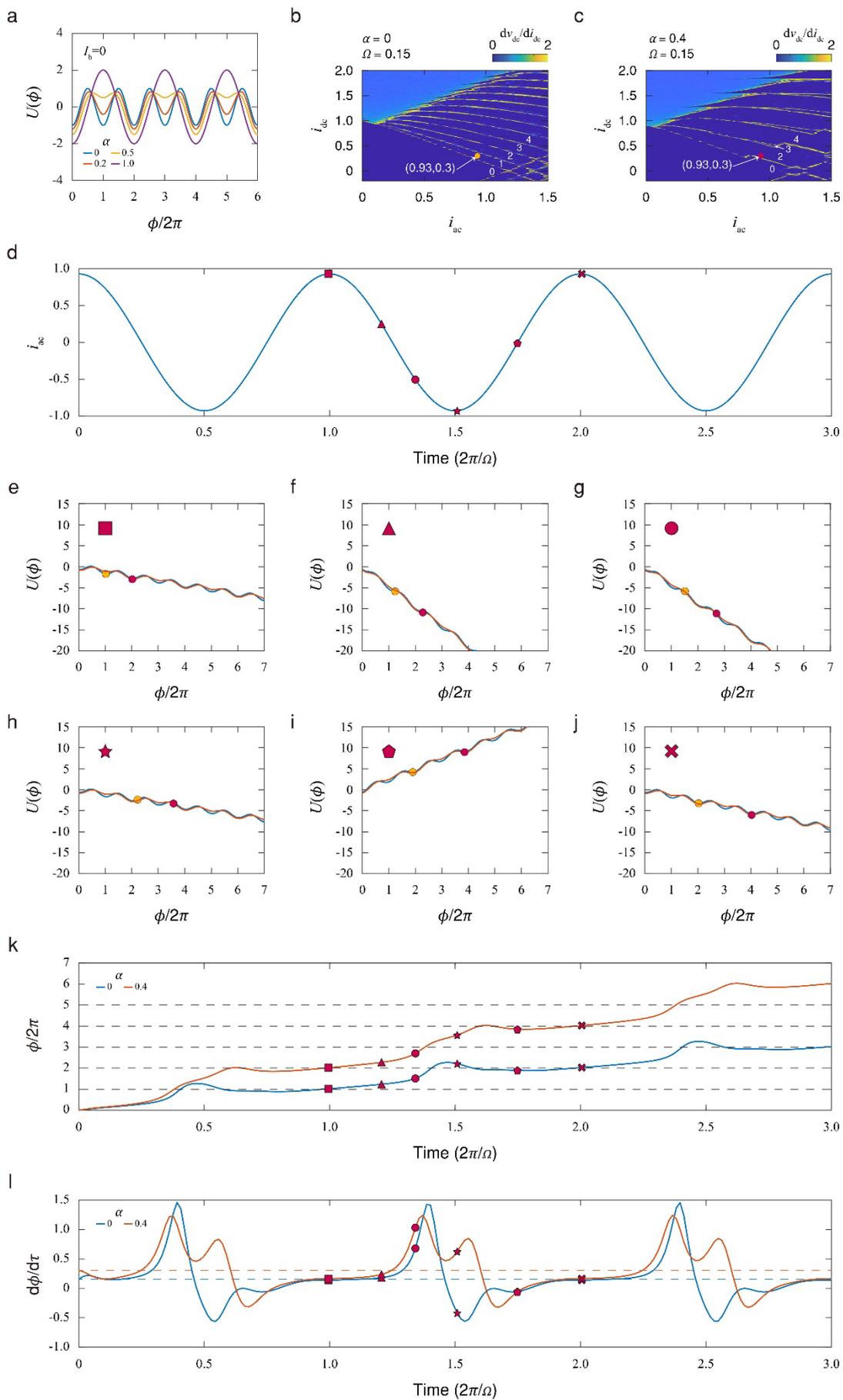

**Fig. S7. Analysis of Shapiro Steps and Washboard Potentials. a**, The washboard potential $U$ as a function of phase difference $\phi$ at various values of $\alpha$ for bias current $I_b = 0$. **b, c,** Numerically calculated Shapiro steps for $\beta=1.6$ as a function of $i_{ac}$ and bias current $i_{dc}$ with varying $\alpha$ and dimensionless frequency $\Omega = hf/2eI_cR$. Here, $\alpha$ represents the ratio of the $4\pi$-periodic Josephson current to $2\pi$-periodic one, $i_{ac}$ and $i_{dc}$ are denoted normalized ac and dc current, respectively. Numbers represent $n$-th voltage steps. **d**, ac current $i_{ac}$ as a function of reduced time $\tau$. Shapes correspond to specific time points represented in the subsequent washboard potentials (**e–j**). **e-j**, The washboard potentials at specific time points marked by symbols in (**d**). Red(yellow) circle represents the phase particle in the case of $\alpha = 0.4$ ($\alpha = 0$) . **k, l,** Phase (**k**) and phase velocity $d\phi/d\tau$ (**l**) as a function of reduced time $\tau$ for $\alpha = 0$ (blue) and $\alpha = 0.4$ (orange). Dotted horizontal lines in (l) represent average values of $d\phi/dt$.

As the frequency $\Omega$ increases, the oscillation period of washboard potential become shorter. This makes it less likely for phase particles to overcome the second hills at low ac and dc current. As a result, $Q_{12}$ increases as $\Omega$ increases. Finally, $\beta$ is inverse proportional to the damping coefficient. A larger $\beta$ means low damping, where the phase particle more likely escapes the potential wells. This result in overall decreases of $Q_{12}$ for given $\alpha$ and $\Omega$.

## S8. $Q_{12}$ dependence in other junctions

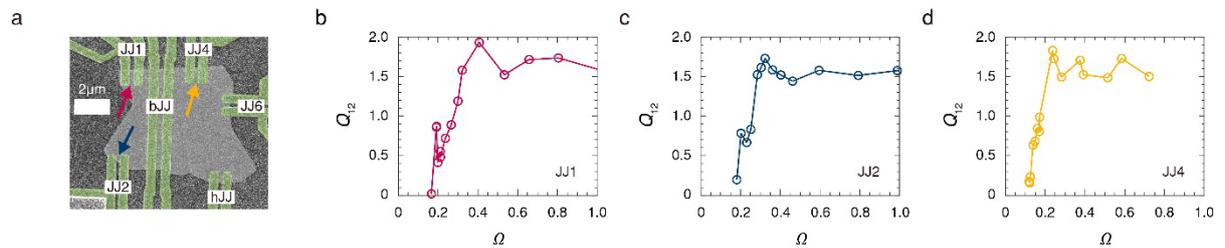

**Fig. S8. Microwave frequency dependence of $Q_{12}$ of other junctions. a**, False-colour scanning electron micrograph. **b-d**, The ratio of $Q_{12}$ for JJ1, JJ2 and JJ4, respectively, as a function of dimensionless frequency, $\Omega = hf/2eI_cR$. The arrow colours in (**a**) correspond to the colours of the curves.